%% file: VLQnlo_skeleton.tex
\documentclass[aps,twocolumn,amsmath,amsfonts,nofootinbib,preprintnumbers,
  superscriptaddress,eqsecnum,secnumarabic]{revtex4-1}
\usepackage[utf8]{inputenc}
\usepackage{color,graphicx,hyperref,amsmath,amssymb,multirow,slashed,ulem}
\hypersetup{bookmarks=true,unicode=true,pdftoolbar=true,pdfmenubar=true,
  pdffitwindow=false,pdfstartview={FitH},pdftitle={VLQ},
  pdfauthor={G.~Cacciapaglia, A.~Carvalho, A.~Deandrea, T.~Flacke, B.~Fuks,
    D.~Majumder, L.~Panizzi, H.-S.~Shao},
  pdfsubject={VLQ},
  pdfcreator={Creator},pdfproducer={Producer}, pdfkeywords={k1}{k2}{k3},
  pdfnewwindow=true,colorlinks=true,
  linkcolor=blue,citecolor=magenta,filecolor=magenta,urlcolor=cyan}

\begin{document}

\newcommand{\be}{\begin{equation}}
\newcommand{\ee}{\end{equation}}
\def\bsp#1\esp{\begin{split}#1\end{split}}

\title{Next-to-leading-order predictions for single vector-like quark production
at the LHC}

\author{Giacomo Cacciapaglia}
\affiliation{Univ Lyon, Universit\'e Lyon 1, CNRS/IN2P3, IPNL, 69622 Villeurbanne, France}
\author{Alexandra Carvalho}
\affiliation{NICPB, Akadeemia tee, Tallinn, Estonia}

\author{Aldo Deandrea}
\affiliation{Univ Lyon, Universit\'e Lyon 1, CNRS/IN2P3, IPNL, 69622 Villeurbanne, France}

\author{Thomas Flacke}
\affiliation{Center for Theoretical Physics of the Universe, Institute for
   Basic Science (IBS), Daejeon 34126 Korea}

\author{Benjamin Fuks}
\affiliation{Institut Universitaire de France, 103 boulevard Saint-Michel,
   75005 Paris, France}
\affiliation{Laboratoire de Physique Th\'eorique et Hautes Energies (LPTHE),
  UMR 7589, Sorbonne Universit\'e et CNRS, 4 place Jussieu,
  75252 Paris Cedex 05, France}

\author{Devdatta Majumder}
\affiliation{The University of Kansas, Lawrence, USA}

\author{Luca Panizzi}
\affiliation{Department of Physics and Astronomy,
Uppsala University, Box 516, SE-751 20 Uppsala, Sweden}
\affiliation{School of Physics and Astronomy, University of Southampton,
  Highfield, Southampton SO17 1BJ, UK}

\author{Hua-Sheng Shao}
\affiliation{Laboratoire de Physique Th\'eorique et Hautes Energies (LPTHE),
  UMR 7589, Sorbonne Universit\'e et CNRS, 4 place Jussieu,
  75252 Paris Cedex 05, France}

\date{\today}

\begin{abstract}
We propose simulation strategies for single production of third generation vector-like quarks at the LHC, implementing next-to-leading-order
corrections in QCD and studying in detail their effect on cross sections and differential distributions. We also investigate the differences
and the relative incertitudes induced by the use of the Four-Flavour Number
Scheme {\it versus} the Five-Flavour Number Scheme. As a phenomenological
illustration, we concentrate
on the production of vector-like quarks coupling to the third generation of the
Standard Model in association with a jet and assuming standard couplings to
gauge and Higgs bosons.
\end{abstract}

\maketitle

\input{1-intro}
\input{2-model}

\input{3-pheno}

\input{4-conclusion}

\acknowledgements

TF is supported by IBS under the project code IBS-R018-D1, while BF and HSS are
partly supported by French state funds managed by the Agence Nationale de la Recherche (ANR), in the context of the LABEX ILP (ANR-11-IDEX-0004-02, ANR-10-LABX-63).
GC and AD acknowledge partial support from the Labex-LIO (Lyon Institute of Origins) under grant ANR-10-LABX-66 and FRAMA (FR3127, F\'ed\'eration de Recherche ``Andr\'e Marie Amp\`ere'').

\bibliographystyle{utphys}
\bibliography{vlqnlo}
\end{document}

%% file: 1-intro.tex
\section{Introduction}
Many extensions of the Standard Model feature vector-like quarks, or quarks
whose left-handed and right-handed components lie in the same representation of
the Standard Model gauge symmetry group. Their existence is predicted, for instance, in
models featuring extra dimensions, an extended gauge symmetry or a composite
Higgs sector~\cite{Antoniadis:1990ew,Kaplan:1991dc,ArkaniHamed:2002qx,%
Contino:2006qr,Matsedonskyi:2012ym}, and their expected mass scale is generally
such that they could be observed in current and future LHC data. For this
reason, vector-like quark searches play a major role in the ATLAS and CMS
searches for new phenomena. Current searches mostly rely on
signatures induced by both their pair and single production, although present
bounds are mostly driven by the strong pair production mechanism, followed by a
decay pattern in which
each vector-like quark is assumed to decay into a Standard Model gauge or
Higgs boson and a third-generation quark, both legs being potentially decaying
differently. Taking into account an
integrated luminosity of 35--36~fb$^{-1}$, vector-like quarks of about 1.2~TeV
are conservatively excluded, regardless of the decay details~\cite{%
Aaboud:2017zfn,Aaboud:2018wxv,Aaboud:2018pii,Sirunyan:2017pks,Sirunyan:2018omb},
although such a bound can in principle be lowered when exotic decay modes into
non-standard particles exist~\cite{Aguilar-Saavedra:2017giu} or if they decay into 
lighter Standard Model quark generations.

Vector-like quark decays are driven by interactions that also induce
their single (weak) production in hadronic collisions. Although the exact
details of the corresponding production mechanism are model-dependent, single
production becomes more and more relevant with respect to pair production once
the mass of the extra quarks is large enough,
by virtue of a smaller phase-space suppression~\cite{DeSimone:2012fs,%
Matsedonskyi:2014mna,Backovic:2015bca}. Whilst the corresponding backgrounds are
usually large, they can be controlled to a large extent by making use of the
properties of the jets accompanying the singly-produced heavy quark. The
leading jet tends to be forward and can thus be an
efficient handle for background rejection, as illustrated by recent experimental
searches~\cite{Aaboud:2018xuw,ATLAS:2018qxs,Sirunyan:2017ynj,Sirunyan:2018fjh,%
Sirunyan:2018ncp}. All those searches, however, rely on simulations of the
signal at the leading-order (LO) accuracy in QCD, although next-to-leading-order
(NLO) corrections are expected to substantially impact the
predictions~\cite{Campbell:2009gj, Fuks:2016ftf}.

In this work, we study the single production of a vector-like top
partner and investigate the effects of NLO corrections on both the total rates
and differential distributions.
We carefully assess the reduction of the theoretical uncertainties originating
from the inclusion of higher-order contributions, and consider two options for
the number of active quark flavors in the proton, namely four and five. In
the former case, the bottom quark is massive and the associated parton density
is taken vanishing. In the latter case, the bottom quark is massless and all
bottom-quark mass effects are effectively resummed into a bottom parton density.
We detail the consequences of such a choice on the total production rate,
as well as on several exclusive kinematic
observables, such as the final-state jet properties, that play a big role in the
search strategies at the LHC in terms of selection cuts, efficiencies and
background {\it versus} signal discrimination.

The rest of this work is organized as follows. In Section~\ref{sec:model}, we
describe the vector-like quark simplified model that we have adopted for our
study, and provide information on our framework for the precise calculation of physical observables.
Section~\ref{sec:pheno} includes our results and main features of the study. We give our conclusions in
Section~\ref{sec:conclusion}.

%% file: 2-model.tex
\begin{figure*}
  \centering
  \includegraphics[width=0.16\textwidth]{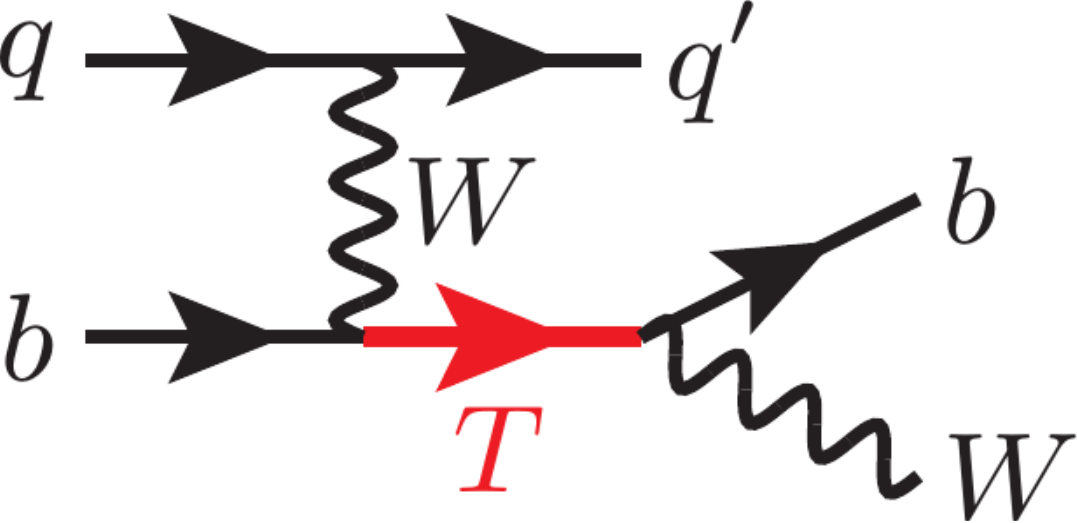}\hspace*{.3cm}
  \includegraphics[width=0.17\textwidth]{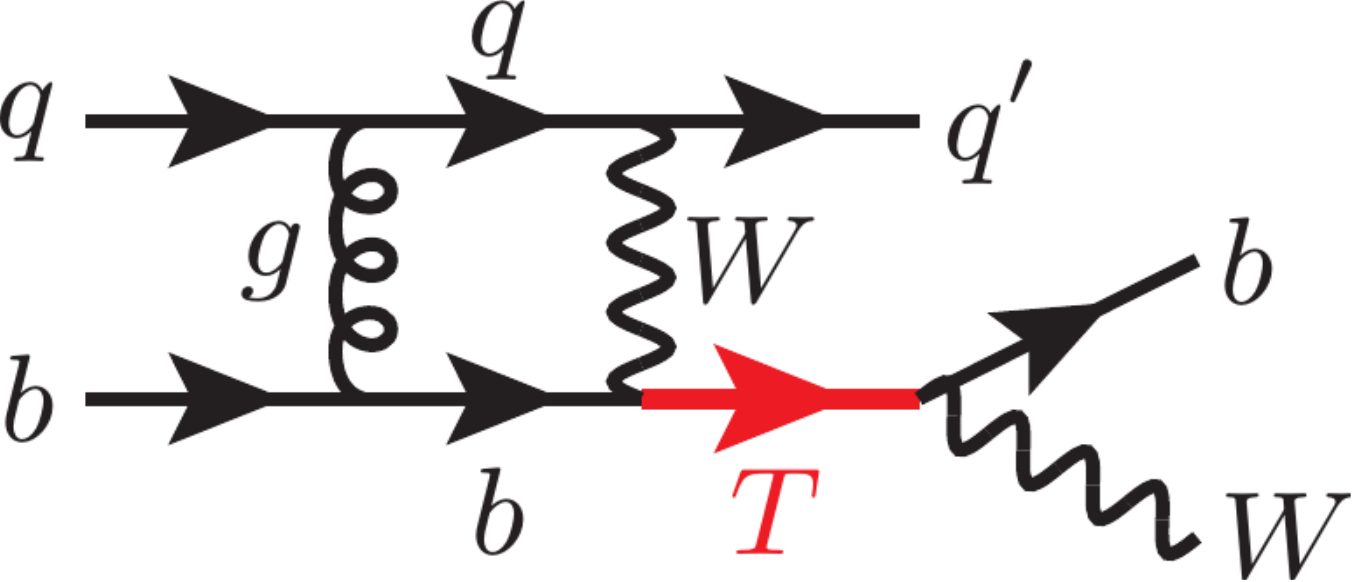}\hspace*{.3cm}
  \includegraphics[width=0.17\textwidth]{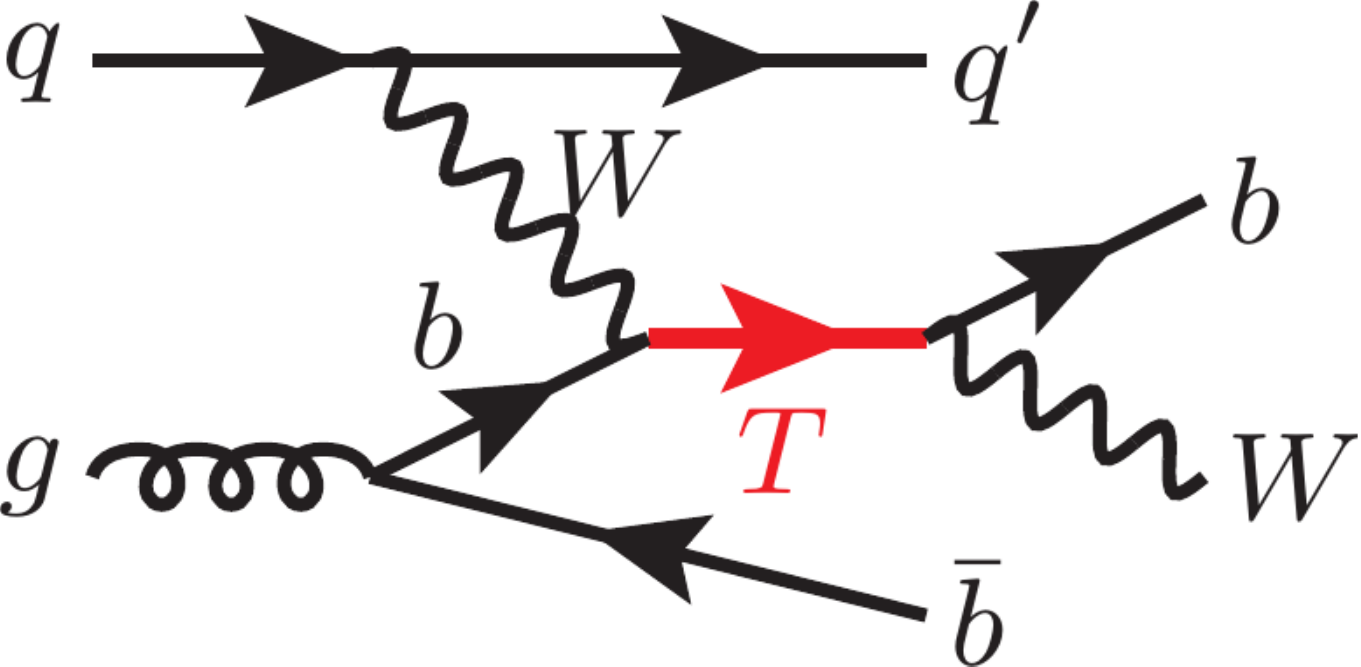}\hspace*{.3cm}
  \includegraphics[width=0.17\textwidth]{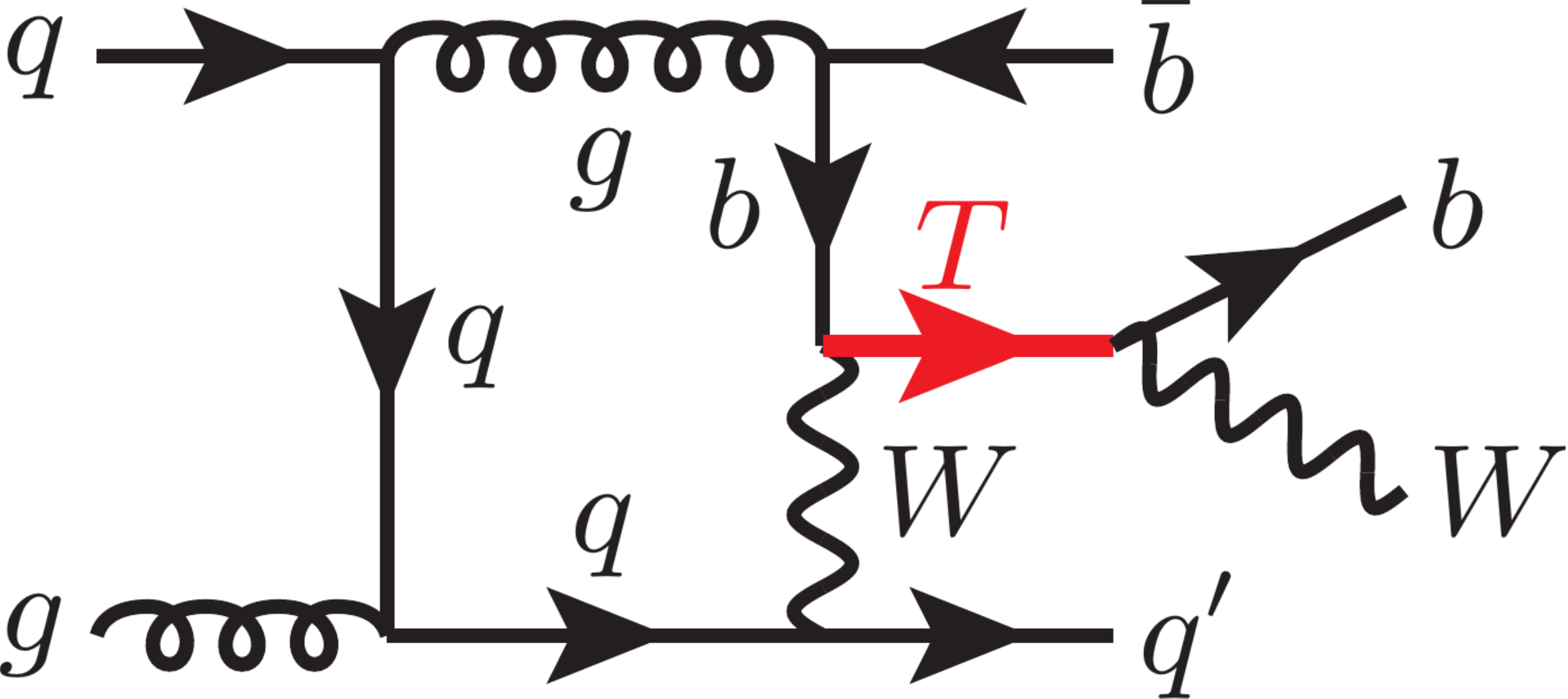}\hspace*{.3cm}
  \includegraphics[width=0.17\textwidth]{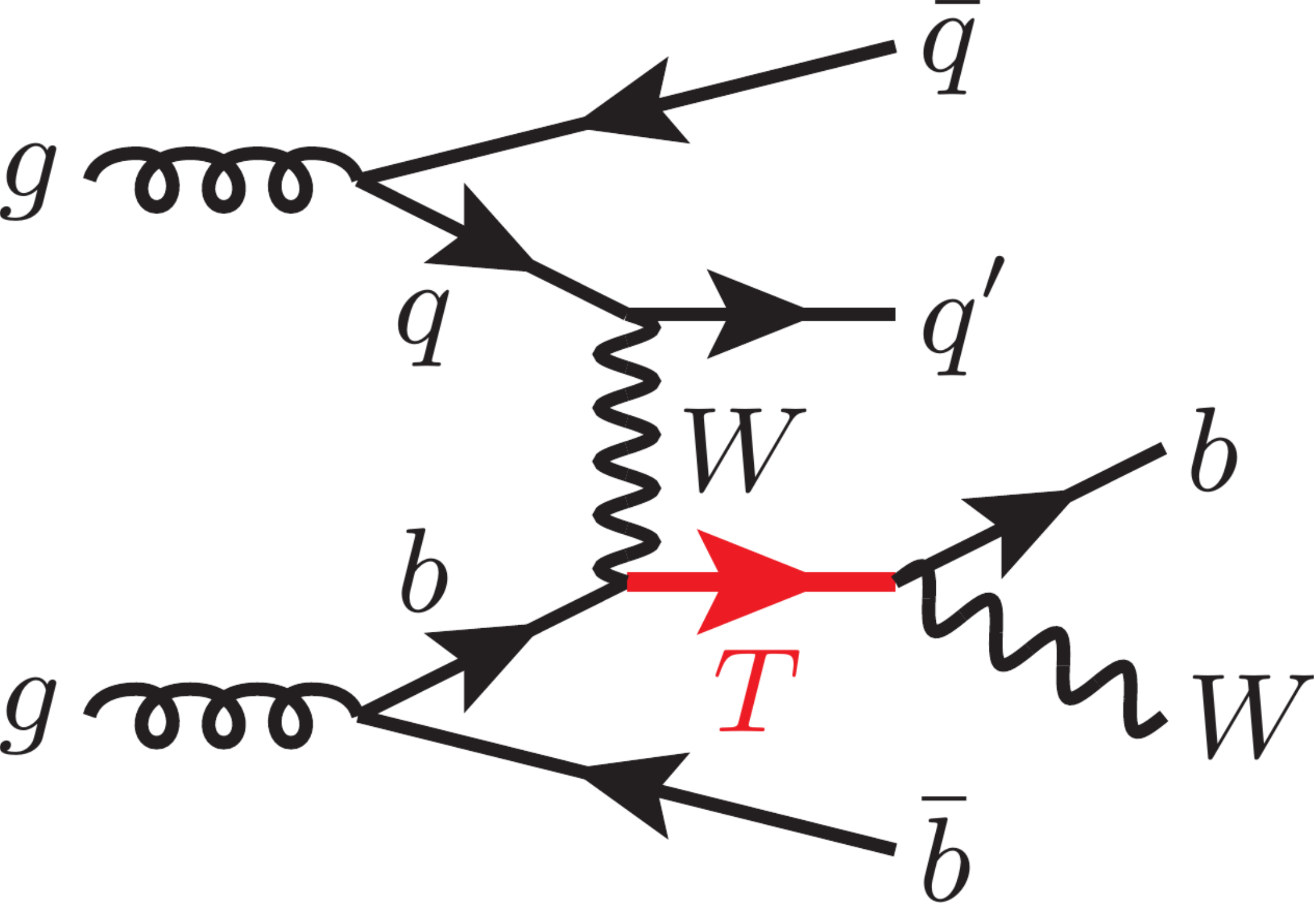}\hspace*{.3cm}
  \caption{\label{fig:topologies}Representative Feynman diagrams for single
    vector-like $T$-quark production. We show typical tree-level (leftmost),
    virtual loop (second leftmost) and real emission (center) contributions in
    the 5FNS, the latter contributing also at the tree-level in the 4FNS.
    Typical 4FNS loop contributions and real emission diagrams are respectively
    presented in the second rightmost and rightmost figures.
   }
\end{figure*}

\section{Theoretical framework and technical details}
\label{sec:model}
In order to investigate the phenomenology associated with the single production
of a vector-like top partner, we consider a simplified model in which the
Standard Model field content is extended by four species of extra quarks $X$,
$T$, $B$ and $Y$ respectively of electric charges 5/3, 2/3, $-1/3$ and $-4/3$.
Focusing on third generation partners, the additional heavy quarks solely couple
with the Standard Model top and bottom quarks, $t$ and $b$, and we forbid any
mixing of the new states with Standard Model lighter quarks. The corresponding
Lagrangian, invariant under $SU(3)_c \times U(1)_Q$ transformations, is then
given by~\cite{Fuks:2016ftf}
\be\bsp
  & \mathcal{L} =
    \sum_{Q=X,T,B,Y} \Big[i \bar{Q} \slashed{D} Q - M_Q \bar{Q} Q\Big] \\
     &\
      - \sum_{Q=T,B}\Big[h\bar Q \big(\hat\kappa_L^Q P_L+\hat\kappa^Q_R P_R\big)
          q + \mathrm{h.c.} \Big] \\
     &\
      + \frac{g}{2 c_W}\sum_{Q=T,B}\Big[ \bar Q \slashed{Z}
          \big( \tilde{\kappa}_L^Q P_L + \tilde{\kappa}^Q_R P_R \big) q +
          \mathrm{h.c.} \Big] \\
     &\
      + \frac{g}{\sqrt{2}} \sum_{Q=X,T,B,Y} \Big[ \bar{Q} \slashed{W}
       \big( \kappa_L^Q P_L \!+\! \kappa_R^Q P_R \big) q \!+\!
        \mathrm{h.c.}\Big]\ ,
\esp\label{eq:LO}\ee
where all the electroweak couplings $\kappa$, $\tilde\kappa$ and $\hat\kappa$ are
considered as free parameters. In the Lagrangian above, $q$ stands for $b$ ($t$) for the neutral-current
interactions of the $B$ ($T$) quark and for the charged-current interactions of
the $Y$ and $T$ ($X$ and $B$) quarks, while $Z$, $W$ are the weak gauge bosons and $h$ is the Higgs boson.
The covariant derivatives include both QCD and QED interactions, while the weak-boson couplings to a pair of
vector-like quarks are omitted as model dependent, usually small and not
relevant for the present study~\cite{Buchkremer:2013bha}. Moreover, $c_W$ stands for
the cosine of the electroweak mixing angle.

We consider the production, at the LHC, of a single vector-like quark and aim to
assess the impact of NLO corrections in QCD on the predictions. To
this aim, we make use of the existing implementation~\cite{Fuks:2016ftf} of the
Lagrangian of Eq.~\eqref{eq:LO} into {\sc FeynRules}~\cite{Alloul:2013bka},
which allows for the generation of an NLO UFO model~\cite{Degrande:2011ua}
through a joint usage of {\sc FeynRules}, NLOCT~\cite{Degrande:2014vpa} and
{\sc FeynArts}~\cite{Hahn:2000kx}. This UFO model contains, in addition to
tree-level model information, ultraviolet counterterms as well as rational $R_2$
Feynman rules~\cite{Ossola:2008xq} so that it can be used within
{\sc MG5\_aMC@NLO}~\cite{Alwall:2014hca} for NLO calculations in QCD. The
cancellation of the ultraviolet divergences appearing in the virtual one-loop
amplitudes is ensured by the presence of the counter-terms in the UFO, and the
finite part of the loop-integrals is evaluated numerically in four dimensions,
the model-dependent and process-dependent parts of the rational terms being
extracted from the $R_2$ information included in the UFO.

Predictions are achieved by convoluting LO and NLO matrix
elements with the NNPDF 3.1 set of parton densities~\cite{Ball:2017nwa}
accessed via the LHAPDF~6 library~\cite{Buckley:2014ana}. After making use of
{\sc MadSpin}~\cite{Artoisenet:2012st} to model the extra quark decays at LO, we match the fixed-order results with the parton shower algorithm of the {\sc Pythia}~8 package~\cite{%
Sjostrand:2014zea}, which we also use, within the
{\sc MG5\_aMC@NLO} package, to simulate the hadronisation process
of all final-state coloured partons. Event reconstruction is
performed with {\sc MadAnalysis}~5~\cite{Conte:2012fm,Conte:2018vmg}, through
which we use the anti-$k_T$ algorithm~\cite{Cacciari:2008gp} with a radius
parameter set to $R=0.4$ as implemented in {\sc FastJet}~\cite{Cacciari:2011ma}.

%% file: 3-pheno.tex
\begin{table*}
 \centering
 \renewcommand{\arraystretch}{1.50}
 \setlength{\tabcolsep}{12pt}
 \begin{tabular}{c | cc| cc}
   \hline
   $M_T$~[GeV] & $\hat\sigma_{\rm LO}^{\rm 4FNS}$ [pb] &
    $\hat\sigma^{\rm 4FNS}_{\rm NLO}$ [pb] & $\hat\sigma_{\rm LO}^{\rm 5FNS}$ [pb] &
    $\hat\sigma_{\rm NLO}^{\rm 5FNS}$ [pb] \\
   \hline
    800  & $32.28^{+27.8\%}_{-20.0\%}{}^{+0.7\%}_{-0.7\%}$ &
           $34.49^{+10.9\%}_{-10.2\%}{}^{+1.4\%}_{-1.4\%}$ &
           $43.75^{+ 2.1\%}_{- 2.7\%}{}^{+1.0\%}_{-1.0\%}$ &
           $41.32^{+ 2.4\%}_{- 1.4\%}{}^{+1.2\%}_{-1.2\%}$ \\
    1200 & $ 8.83^{+32.1\%}_{-22.2\%}{}^{+1.3\%}_{-1.3\%}$ &
           $ 8.59^{+12.8\%}_{-11.7\%}{}^{+1.9\%}_{-1.9\%}$ &
           $11.93^{+ 5.6\%}_{- 5.4\%}{}^{+1.2\%}_{-1.2\%}$ &
           $11.55^{+ 2.2\%}_{- 1.0\%}{}^{+1.6\%}_{-1.6\%}$ \\
    1600 & $ 2.90^{+35.4\%}_{-24.0\%}{}^{+1.5\%}_{-1.5\%}$ &
           $ 2.71^{+13.9\%}_{-12.5\%}{}^{+2.4\%}_{-2.4\%}$ &
           $ 3.87^{+ 8.2\%}_{- 7.3\%}{}^{+1.7\%}_{-1.7\%}$ &
           $ 3.65^{+ 1.9\%}_{- 1.0\%}{}^{+1.9\%}_{-1.9\%}$ \\
    2000 & $ 1.05^{+38.2\%}_{-25.3\%}{}^{+1.9\%}_{-1.9\%}$ &
           $ 0.91^{+15.5\%}_{-13.6\%}{}^{+2.8\%}_{-2.8\%}$ &
           $ 1.40^{+10.4\%}_{- 8.9\%}{}^{+2.7\%}_{-2.7\%}$ &
           $ 1.32^{+ 1.9\%}_{- 0.8\%}{}^{+2.5\%}_{-2.5\%}$ \\
   \hline
  \end{tabular}
  \caption{LO and NLO QCD inclusive cross sections for the single production of
    a heavy $T$-quark in the 4FNS and 5FNS, in the context of LHC collisions at
    a centre-of-mass energy $\sqrt{s}=13$~TeV. The results are shown together
    with the associated scale and parton density uncertainties.
  }
  \label{tab:totalrates}
\end{table*}

\section{Single VLQ production at the LHC}
\label{sec:pheno}
In this section, we investigate the phenomenology associated with the single
production of a vector-like quark
at the LHC, in proton-proton collisions at a centre-of-mass
energy of 13~TeV. We study the impact of the NLO corrections both at the level
of the total rate and for differential cross sections. For the latter, we put
the emphasis on observables crucial for providing handles on the discrimination
of a potential signal from the background, such as the properties of the jets
produced together with the heavy quark.

We consider the production of vector-like quarks interacting with the
Standard Model third-generation quarks, so
that the associated single-production processes involve bottom quarks, as shown
in Fig.~\ref{fig:topologies}. The bottom quarks can be accounted for in the
calculations in two
different ways. In the so-called four-flavour-number scheme (4FNS), bottom
quarks, which are significantly heavier than the proton, are treated as heavy
flavours. Thus,
they cannot contribute to the proton wave-function and are solely produced
as massive final-state particles. In the five-flavour-number scheme (5FNS), the
mass of the bottom quark is neglected with respect to the hard scale of the
processes under consideration, and its effects are resummed into a bottom quark
density in the proton that is not present in the 4FNS. Whilst at all orders in
perturbation theory the two schemes are identical, the way in which the
perturbative expansion is organised is different at a given finite order, so
that predictions may not necessarily match. In this work, we assess the
implications on the single production of a vector-like quark of
the third generation whose dynamics stems from the Lagrangian of
Eq.~\eqref{eq:LO}.

\begin{figure}
  \centering
  \includegraphics[width=.95\columnwidth]{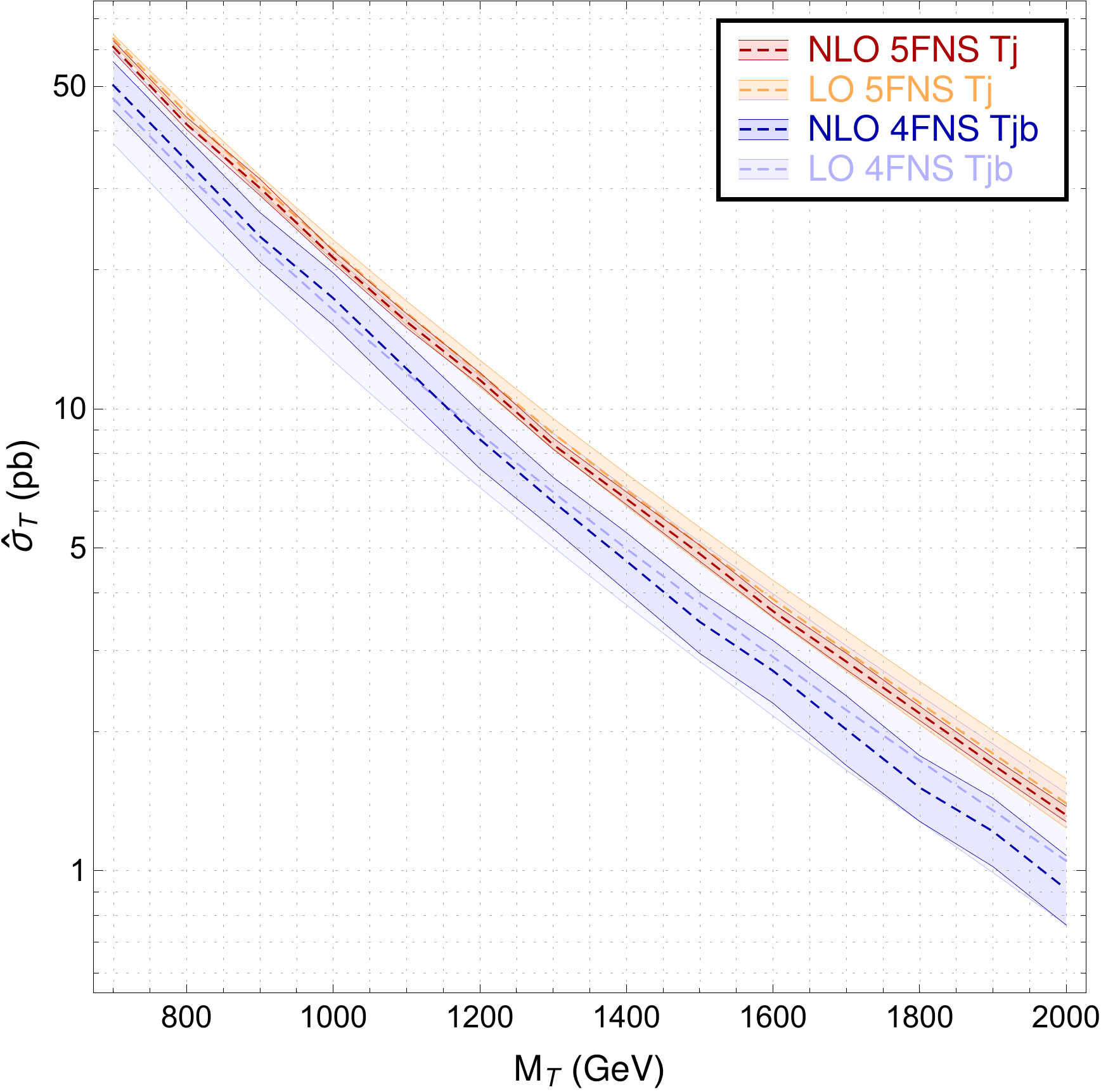}
  \caption{Total production cross section for single $T$-quark production at the
    LHC. We present results at the leading order in the 5FNS (orange) and 4FNS
    (light blue) as well as at the NLO QCD in the 5FNS (red) and in the 4FNS
    (dark blue).
    Scale and parton density 1$\sigma$ uncertainties are included in the error bands.
  }
  \label{fig:totalrates}
\end{figure}

As a benchmark, we consider an up-type vector-like quark $T$ that couples only to the
$W$-boson. 
For simplicity, only the left-handed couplings have been set to non-zero values, 
{\it i.e.} $\kappa^T_L\neq0$ and $\kappa^T_R=0$.
This choice is motivated by the fact that, in
the case of the presence of a single vector-like quark $Q$, only one of the two
mixing angles is large, the other being suppressed by a factor of
$m_q/M_Q$ where $m_Q$ and $m_q$ are the heavy quark and relevant Standard Model
quark masses respectively~\cite{Cacciapaglia:2010vn}.
All our results are given after factorising the whole $\frac{g}{\sqrt{2}}\kappa^T_L$ parameter, 
such that the total production cross-section $\sigma_T$ is equal to $(\frac{g}{\sqrt{2}}\kappa^T_L)^2\hat\sigma_T$.

In the 5FNS, single $T$-quark production originates from a $2\to2$ process where
the heavy quark is produced in association with a jet that is mainly forward, as
illustrated in the leftmost diagram of Figure~\ref{fig:topologies}. The
$T$-decay into a $Wb$ system, {\it i.e.} the only allowed decay for the
considered scenario, is also shown. Typical loop and real emission contributions
are presented in the second and third diagram of the same figure. In the 4FNS,
there is no $b$-quark density in the proton so that, at the lowest order, the
heavy quark can only be produced after the splitting of a gluon into a
bottom--anti-bottom pair, as illustrated by the third diagram of
Figure~\ref{fig:topologies}. NLO contributions to this process are illustrated
by the fourth (virtual loop) and fifth (real emission) diagrams of the figure.
Whilst some diagrams are common to both schemes, we recall that the mass of the
bottom quark is neglected in the 5FNS whilst kept non-vanishing in the 4FNS, on
top of the presence of a bottom density in the 5FNS.

\begin{figure*}
  \centering
  \includegraphics[width=0.4\textwidth]{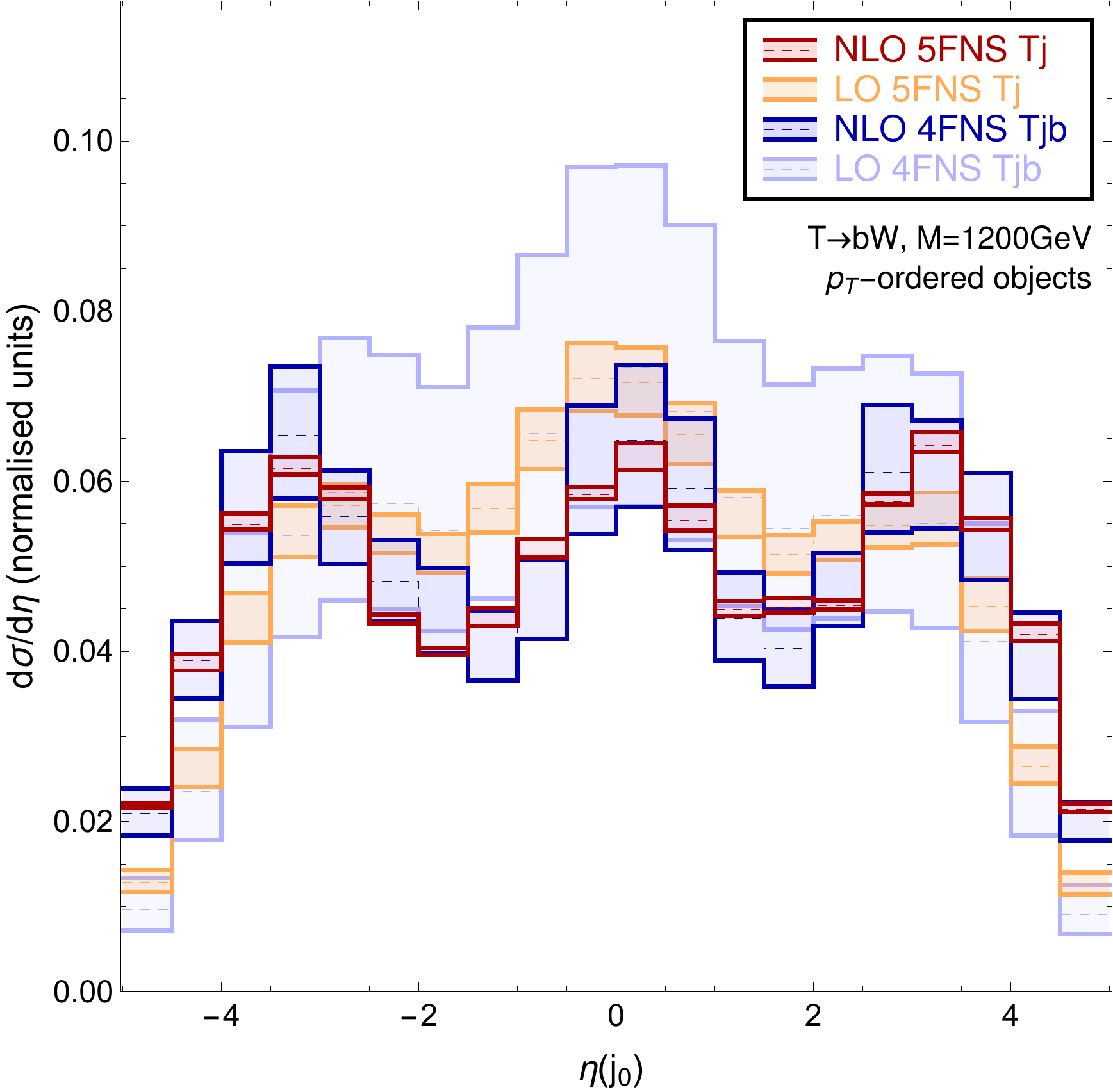}\hskip 10pt
  \includegraphics[width=0.4\textwidth]{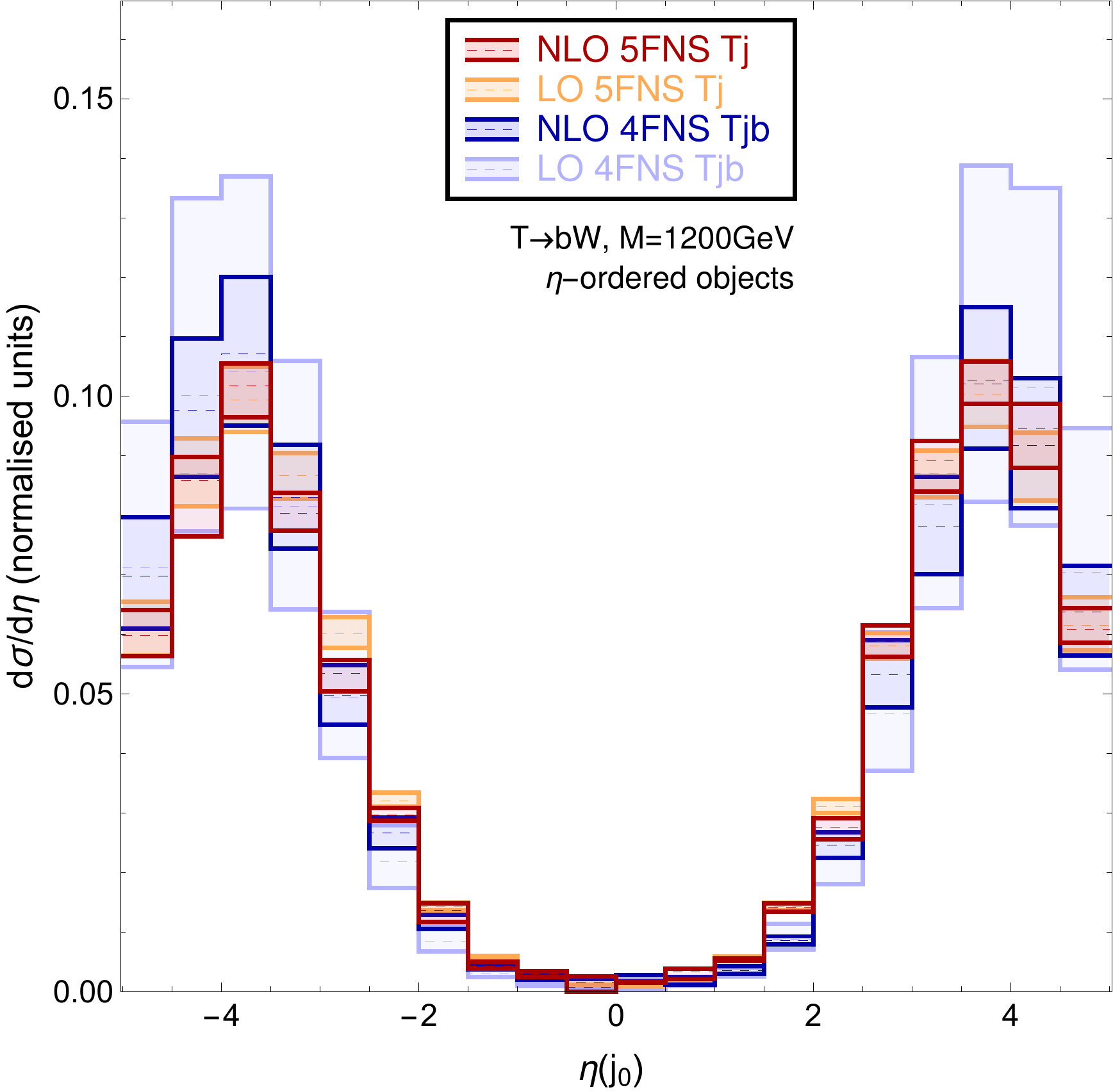}
  \caption{Normalised distributions in the pseudorapidity of the
    leading non-$b$-jet, when the jets are ordered by transverse momentum (left)
    and absolute value of pseudorapidity (right). Results are given both
    at the LO and NLO accuracy and for the 4FNS and 5FNS, adding linearly
    scale and parton density uncertainties in the error bands.}
  \label{fig:diffrate_eta}
\end{figure*}

In Figure~\ref{fig:totalrates} and Table~\ref{tab:totalrates}, we present LO and
NLO cross section values for the single production of a vector-like $T$-quark,
both in the 4FNS and 5FNS, and illustrate the dependence of the results on the
vector-like quark mass $M_T$. The results are presented together with their
scale uncertainties, evaluated by independently varying the renormalisation and
factorisation scales by a factor of 2 up and down relatively to a central scale
set to half the scalar sum of the transverse momenta of all final state
particles, as well as with
their parton density uncertainties extracted following the recommendations of
Ref.~\cite{Demartin:2010er}. Whilst LO parton densities should in principle be
used when a convolution with LO matrix elements is in order, we have instead
made use
of the NLO NNPDF 3.1 set in all calculations due to the poor quality of the LO
fit~\cite{Ball:2017nwa}. In the figure, the two contributions to the total
error are added linearly, whereas we show them separately in the table.

In the 4FNS, NLO corrections increase the central value of the cross
section by about 10\% in the low-mass region, whilst
reducing it by about 10\% in the higher mass regime, the NLO-to-LO cross section
ratio being about 1 for a heavy quark mass of about 1.1~TeV. In contrast, in
the 5FNS, NLO effects always affect the cross section by reducing it by about
10\%, regardless of the actual $M_T$ value. This reduction of the cross section
is connected to the virtual amplitudes appearing at NLO and contributing
significantly.
These channels are illustrated by the second diagram of
Figure~\ref{fig:topologies} (in the 5FNS) and the fourth diagram of
Figure~\ref{fig:topologies} (in the 4FNS).

Whilst results in the two schemes largely differ at LO, the compatibility
between the predictions is expected to improve when higher orders are included.
In the small mass regime, the compatibility between the 4FNS and 5FNS
perturbative expansions is indeed improved at NLO, the differences between the
total rates being of about 20\% (to be compared with a 30\% difference at LO).
However, for heavy quark masses $M_T$ larger than 1.1~TeV, this is not
the case anymore and the difference between the two schemes stays constant at
about 30\%, both at LO and NLO. 
This is due to the fact that, whilst finite bottom mass power corrections of the type
$(m_b^2/Q^2)^n$ are suppressed and negligible for the considered high scale
$Q$, logarithms $\log Q^2/m_b^2$ could be large and may impact the two
considered perturbative expansions. In this case, higher-order corrections, or
the all-order resummation included in the 5FNS, could be necessary to get
cross section estimates that agree better with each other~\cite{Maltoni:2012pa}.
Differences between the results obtained in the 4FNS and 5FNS persist at NLO for
vector-like quarks with mass above about 1~TeV, so that this corresponds to a
parameter space region in which large logarithms of the bottom quark mass play a
significant role. In contrast, the results are perturbatively well-behaved in
the light $T$-quark case, and a substantial agreement can be found between the
predictions in both schemes. Finally, a significant reduction of
the scale uncertainties is obtained in both schemes at NLO, the improvement
reaching up to 50\% compared with the respective LO results. Total
rates however genuinely feature smaller uncertainties
in the 5FNS, as stemming from the resummation of the logarithms in the bottom
quark mass. As predictions matching both schemes consistently are
not available, the 5FNS total cross section results seem preferrable due to the
smaller associated uncertainties and more stable $K$-factor (defined as the
ratio of the NLO to LO predictions) with respect to $M_T$ variations.

In the following, we asses the impact of the NLO corrections in both
schemes by considering more exclusive observables, like those related to the
final-state jet properties. We shall focus on the kinematics of the accompanying
light jet produced in association with the vector-like quark, as well as on the
$b$-jet originating from the vector-like-quark decay. As a benchmark, we
fix the heavy quark mass to $M_T = 1.2$~TeV, a value close to the average
current exclusion bounds~\cite{Aaboud:2017zfn,Aaboud:2018wxv,Aaboud:2018pii,%
Sirunyan:2017pks,Sirunyan:2018omb}. We recall that, since the only non-vanishing
new physics
coupling in Eq.~\eqref{eq:LO} is $\kappa_L^T$, the heavy
quark $T$ can only decay into a $W$-boson and a $b$-jet. We moreover enforce the
$W$-boson to decay muonically for simplicity.

\begin{figure*}
  \centering
  \includegraphics[width=0.33\textwidth]{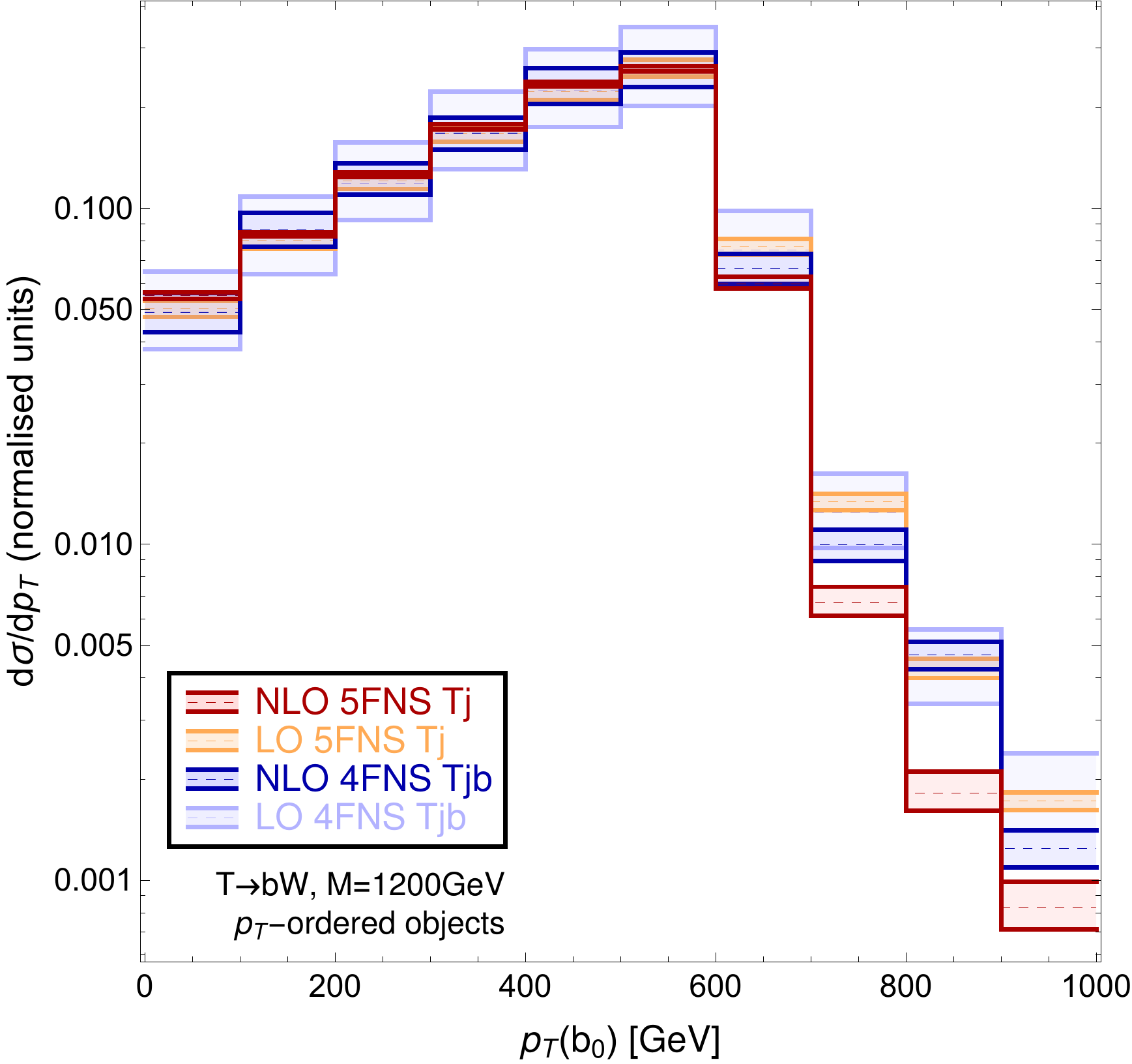}\hfill
  \includegraphics[width=0.315\textwidth]{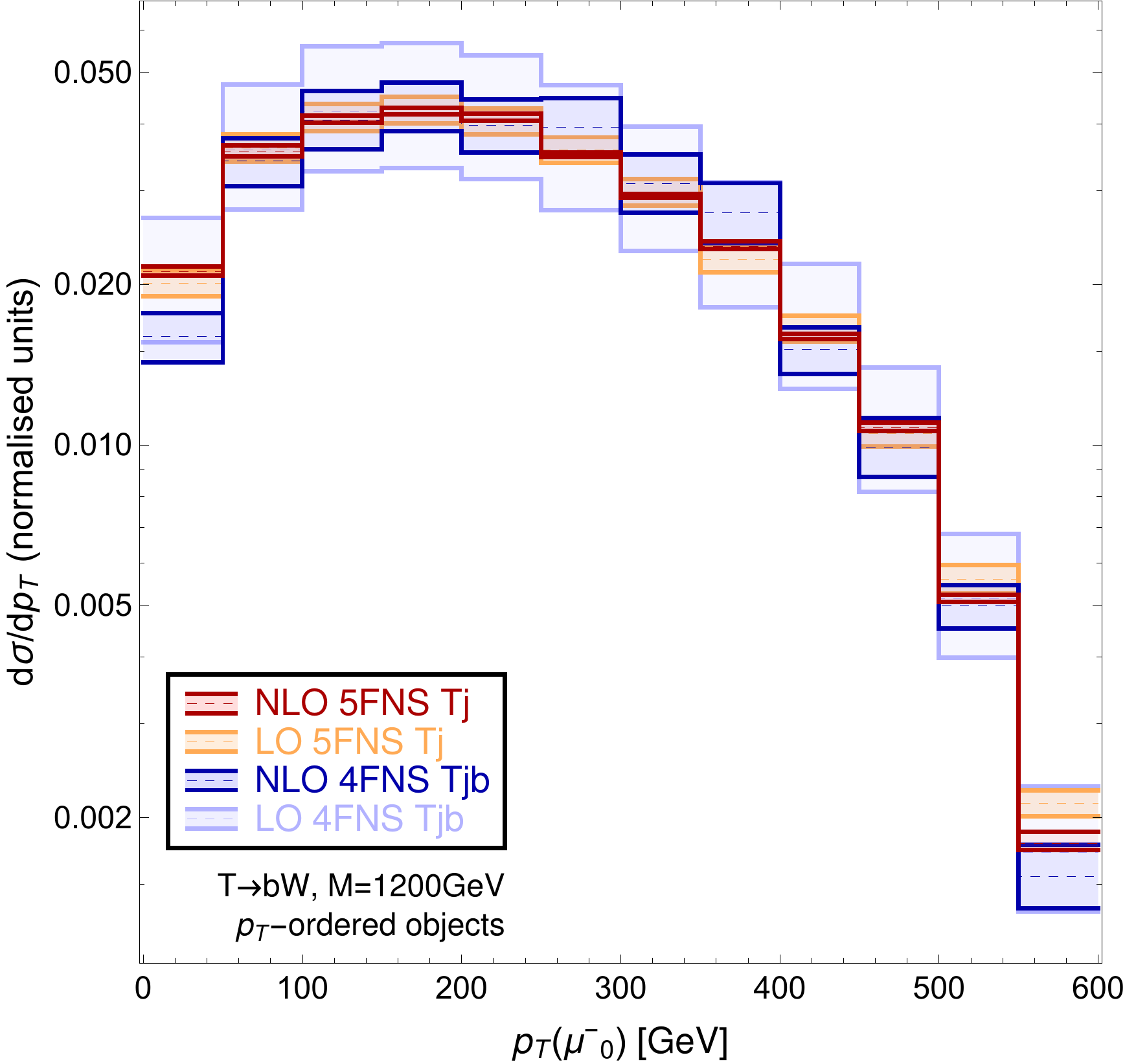}\hfill
  \includegraphics[width=0.33\textwidth]{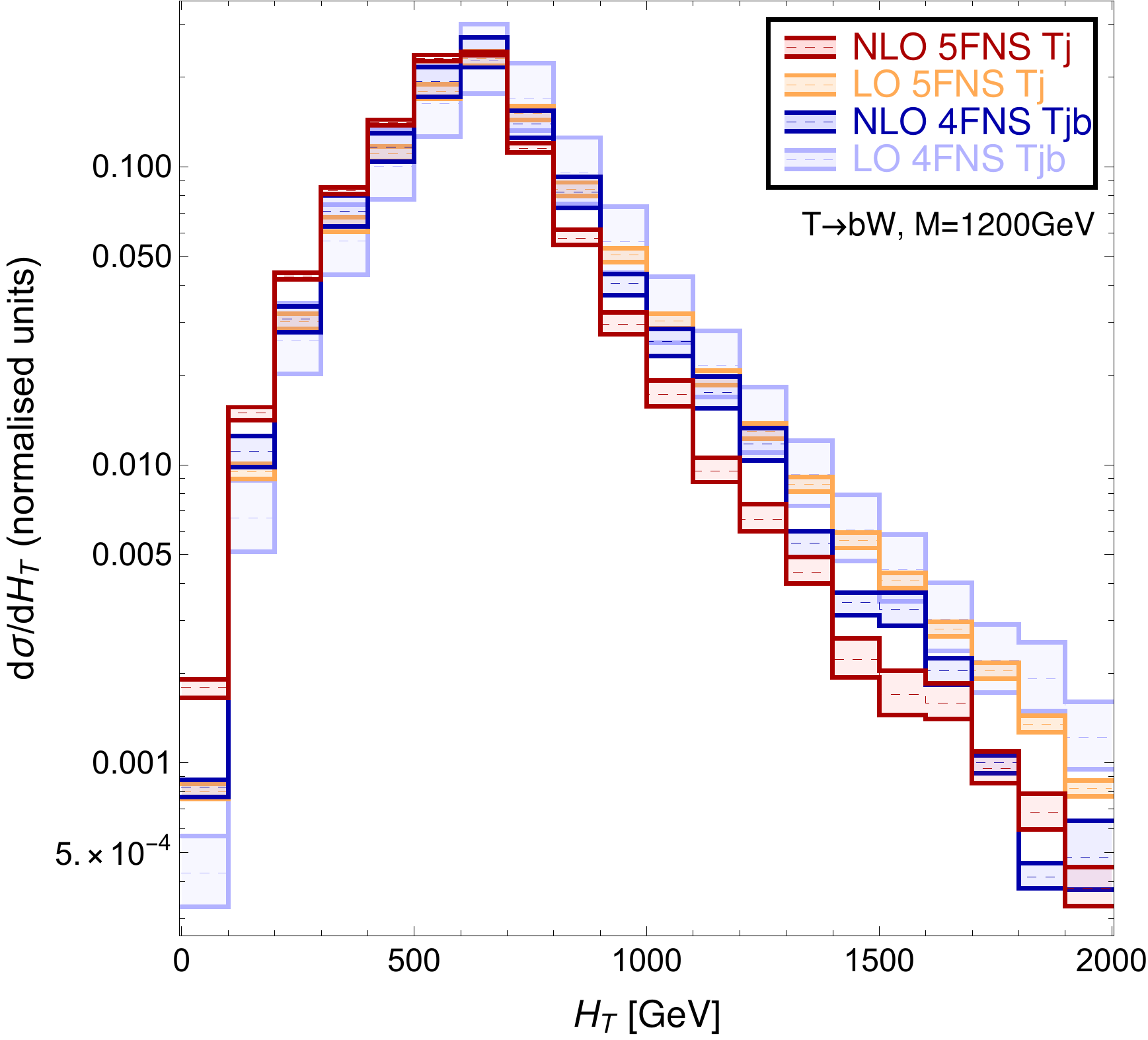}
  \caption{Normalised distributions in the transverse momentum of the leading
    $b$-jet (left) and of the leading muon (centre) both for $p_T$-ordered
    objects, as well as in the $H_T$ variable
    representing the hadronic activity (right). We present results both
    at the LO and NLO accuracy and for the 4FNS and 5FNS, adding linearly
    scale and parton density uncertainties in the error bands.}
  \label{fig:diffrate_pt}
\end{figure*}

In Figure~\ref{fig:diffrate_eta}, we present the
pseudorapidity ($\eta$) distributions of the leading non-$b$-jet $j_0$,
{\it i.e.} the leading jet for which there is no $B$-hadron lying in a cone of
radius $R=0.4$ centred on the direction of the jet. We show the distributions
in two cases, first when the objects are ordered according to their transverse
momentum $p_T$ (left panel) and second when they are ordered according to the
absolute value of their pseudorapidity (right panel). As already mentioned, this
jet is of particular importance for the searches for vector-like-quark single production,
as it is preferably produced in the forward direction by virtue of the
$t$-channel production mode of a single massive particle. This forward character
is indeed crucially used by the experimental searches as a handle for unraveling
the signal from the background for which the corresponding events feature more
central jets~\cite{Aguilar-Saavedra:2013qpa}.
As shown on the left panel of the figure, the hardest jet (in transverse
momentum) is indeed mostly forward, the distribution featuring
local maxima around $|\eta| \sim 3$, and this hardest jet is often the most
forward one as demonstrated by the results in the right panel of the figure.
Moreover, the inclusion of the NLO
contributions is crucial to get the theoretical uncertainties under good
control, as the LO predictions are plagued by uncertainties
larger than 30\%, a fact that can be seen as a loss of predictive power.
Focusing on the NLO results, 4FNS (dark blue) and 5FNS (dark red) predictions
reasonably agree with each other, showing that the importance of the logarithms
in the bottom quark mass is mild for what concerns the shape of the
differential distributions. Their resummation indeed only affects the total
rate, as shown in Figure~\ref{fig:totalrates}. However, the bin-by-bin NLO-to-LO
ratio is far from being constant. Whereas in the case of pseudorapidity-ordered
objects the shapes of the distributions are barely affected by the NLO
corrections, the situation changes when the objects are $p_T$-ordered, so that
NLO distributions are necessary for a precise determination of the observable.

In Figure~\ref{fig:diffrate_pt}, we investigate other
properties of single vector-like quark production, like the
transverse momentum spectrum of the leading $b$-jet (left panel) that
usually originates from the heavy quark decay, as well as the one of the leading
muon (central panel) that is issued from the decay of the $W$-boson stemming
from the heavy quark decay. As all particles can be seen as essentially massless
compared with the mass scale of the heavy quark $T$,
we expect these two distributions to respectively peak around $M_T/2$ and
$M_T/4$, as driven by the structure of the heavy quark decay,
\be
  T \to W  b \to \mu \nu_\mu b \ .
\ee
The available energy is equally shared between the $b$-jet and the
$W$-boson, such that both get a transverse momentum of ${\cal O}(M_T/2)$. The
decay products of the $W$-boson are then expected to feature a $p_T$
distribution peaking at about $M_T/4$. This property could be additionally
illustrated by
the missing transverse-momentum spectrum that would exhibit a similar structure
as the muon $p_T$ spectrum. The relevance of the NLO corrections can be seen
especially in the tail of the leading $b$-jet $p_T$ distribution, where the NLO
distributions have a tendency to fall steeper with respect to their LO
counterparts by virtue of the larger impact of the new channels for these phase
space configurations.

In the right panel of the figure, we
describe the hadronic activity typically associated with single vector-like
quark production by presenting the $H_T$ spectrum, where the $H_T$ variable is
defined as the scalar sum of the transverse momentum of all reconstructed jets
of the event. The peak at about $M_T/2$ is due, once again, to the $W$-boson
that carries away half of the available energy through its leptonic decay, and
that is then not accounted for in the computation of the $H_T$ variable. Also in
this case the NLO distributions exhibit a steeper fall as a function of $H_T$,
with respect to the LO ones. Nevertheless, the shape of the $H_T$ observable may
be impacted by contributions originating from higher orders. The bulk of the
effect could however be derived by implementing the merging of multipartonic
matrix elements featuring different jet multiplicities, both at LO and NLO. Such
a study however lies beyond the scope of this work.

To summarize, in general we observe a substantial reduction of the uncertainties at NLO, and
the 4FNS and 5FNS usually agreeing (in shape) with each other.

%% file: 4-conclusion.tex
\section{Conclusions}
\label{sec:conclusion}

We have studied the phenomenology associated with the single production of
a vector-like quark that couples to the third generation quarks of the Standard
Model and to the weak gauge and Higgs bosons.
We have estimated the impact of the NLO corrections in QCD, both in terms of the
total production rate and of the differential distributions for the illustrative
example of an up-type top partner.  In the
4FNS, we have observed that the NLO-to-LO ratio of
the total rate strongly depends on the vector-like quark mass, moving from values larger than 1
to values smaller than 1 when going from lower to higher masses. In
the 5FNS, the same ratio is more stable, being always of the order of 0.95 in the whole mass range. Moreover, the
shapes of the  distributions are, to a large extent, independent of the way in
which the bottom quark mass is treated and thus on the number of active quark
flavours. We have also discussed in detail the features present in the distributions
used for the experimental searches (both in the 4FNS and in the 5FNS) and in particular the role of the leading
jets in these distributions.
Regardless of the scheme used for the calculations, NLO corrections are
mandatory for getting the theoretical uncertainties under control.

Our results therefore demonstrate that the inclusion of the NLO corrections in
QCD are essential in order to properly study the single production of
vector-like quarks at hadron colliders, like the LHC, and that the allowed
different treatments of the bottom quark only affect the rates and the
distributions in specific corners of the phase space.